\documentstyle[aps,manuscript]{revtex}  % DON'T CHANGE
\begin{document}
\draft
\title { The modifications of customary filtrational equation }

\author{ M.N.Ovchinnikov}

\address{Faculty of Physics, University of Kazan,
Kremlevskaya 18, 420008, Kazan, Russia, Marat.Ovchinnikov@ksu.ru }

\date{\today}

\maketitle

\begin{abstract}
{\it 
The usable limits of the customary and relaxational   filtrational theories are
considered. The questions of applicable  the locality  and local 
thermodynamical equilibrium principles to depict the nonstationary flows are 
discussed. The experimental procedures are proposed to determine the 
filtrational flows relaxation times.}
\end{abstract}


\narrowtext

The theoretical and experimental investigations of the filtrational processes 
in porous media makes for a long time and the complexity of such systems does 
not enables to descript their evolution in a simple manner. 
The nonlinear effects are an essetial in some situations. The locality and
local thermodynamical equilibrium principles applicability remains to be
investigated also. Below we will be take into consideration the linear 
theories only.

The Darcy equation

\begin{equation}
     \vec W=-\frac{k}{\mu} \nabla P
\end{equation}

was obtain from experiments under stationary filtration conditions. 
To descript the nonstationary processes usually used the continuously and 
the state equations in form

\begin{equation}
     m(P)=m_{0}+\beta_{m} (P-P_{0})
\end{equation}
\begin{equation}
     \rho (P)=\rho_{0}(1+\beta_{f} (P-P_{0}))
\end{equation}
\begin{equation}
    \frac{\partial (m \rho)}{\partial t}+div(\rho \vec W)=0
\end{equation}

Now can produced the customary filtrational equation now as

\begin{equation}
     \frac{\partial P}{\partial t}-\ae \Delta P=0
\end{equation}
where  {\ae} - piezoconductivity, k - permeability, $\mu$ - viscosity, 
P - pressure, $\vec W$ - filtration velosity  , $\rho$ - fluid density,
$m$ - porosity, $\beta_{m}$ and $\beta_{f}$ - compressibility of porous
matrix and fluid respectively.

The fundamental solution of (5) for onedimesional system is

\begin{equation}
P(x,t)=\frac{\Theta (t)}{\sqrt {4 \ae \pi t}} exp( -\frac{x^2}{4  \ae t}   )
\end{equation}

where $\Theta(t)$ is the Heaviside function.
 
We can see from (6) that the customary filtrational equation leads to infinity 
phase and group velosities paradox  like equations for classical heat 
conductivity and diffusion.

It should be mentioned that the questions of locality  and local 
thermodynamical equilibrium principles applicability for the systems under 
investigation are dicussed seldom[1]. In this aspect let us assume the 
solution of equation (5) for the case of plane  parallel onedimensional
filtration with the constant pressure difference
by the frontiers ($P_{f}$)

\begin{equation}
P(x,t)=P_{f} (1-x/L-\sum_{n=0}^{\infty}(2/(\pi n)) Sin(\pi n x/L)
 exp(-\pi^2 n^2 \ae t/L^2))
\end{equation}

The multiexponential dependences pressure from time make it possible to
introduce the characteristic time 
of the transition to the stationary state as $\tau^{\ast} =0.1*L^2/\ae$, 
where L- the distance between frontiers.
Now we can estimate this times. Let  L=100 meters,  $\ae=1m^2/sec$, 
then $\tau^{\ast} \sim 10^3 sec$.  
If  L=1 m, $\ae=1m^2/sec$, $\tau^{\ast} \sim 0.1$,  when
$L<10^{-2}$, $\ae=1m^2/sec$, $\tau^{\ast}<10^{-5} sec$, and in the last case
we have a situation when the velosity of stationary state establising becomes
more than the sound velosity in this media. It is a strange conclution. Where
is the time and space usable limits of the filtrational theories?

One of the effective attempt to resolve this situation is the relaxational
theory [2]. This theory takes into account that the local equilibriun is 
established in time with the according the next relaxational equation 
($\tau_{w}$ - time of relaxation)

\begin{equation}
     \vec W+\tau_w\frac{\partial \vec W}{\partial t} =-\frac{k}{\mu} \nabla P
\end{equation}

Actually this is the local nonequilibrium procedure. In according (8)  we come
to the hiperbolic equation

\begin{equation}
   \frac{\partial P}{\partial t}+\tau_w\frac{\partial^2 }{\partial t^2}P -\ae 
\Delta P=0
\end{equation}

with the finite phase and group velocities  $( V_{ph} = \sqrt(\ae/\tau) )$. 

In some cases author written relaxation equation in the double relaxational 
form

\begin{equation}
     \vec W+\tau_w\frac{\partial \vec W}{\partial t} =-\frac{k}{\mu}
 \nabla (P+\tau_P\frac{\partial  P}{\partial t} )
\end{equation}

and in that event we returns to parabolic form the filtrational equation 
with the infinite group and phase velosities

\begin{equation}
     \frac{\partial P}{\partial t}+\tau_w\frac{\partial^2}{\partial t^2}P
-\ae \Delta (P+\tau_p\frac{\partial P}{\partial t}) =0
\end{equation}

\begin{equation}
    V_{ph}=Re(\sqrt{ \omega  \ae}\sqrt{\frac{1+i\omega\tau_p}
{-i+\omega\tau_w}})
\end{equation}

To test the validity the relaxational filtration theory we may to carry out
the experiments with so calles filtrational waves, when the harmonic 
oscillations of pressure is created in porous media. In case $\omega\tau<<1$ 
the relatation theory leads to the declination $\sim \omega\tau/2$  in phase 
velosities relative to customary equations (5).
But in high frequences the relaxation theory have to tends to Biot [3] theory
for waves in saturated porous media. So, it is nesessary to explore the 
investigations in this area. For instance we can investigate this process by
means of molecular dynamics simulation [4] and produce the filtrational
law averiging the Navier and Stokes equation [5].

\newpage

\begin{enumerate}

\item  Sobolev S. L. Phys.Rev. E, 55 4, 1997.
\item  Molokovich Yu.M. Izvestia vuzov. Mathematics, 1977, 8, pp.49-55.
\item  Biot M.A. J.Acoust.Soc.Amer., vol.88, 1956, pp.168-186.
\item  Ovchinnikov M.N. (in press).
\item  Sanchez-Palencia E. Int.Jour.Eng.Sci., 1974, 12, pp.331-351.

\end{enumerate}

\end{document}